\documentclass[pdflatex,sn-nature]{sn-jnl}

\usepackage{graphicx}%
\usepackage{multirow}%
\usepackage{amsmath,amssymb,amsfonts}%
\usepackage{amsthm}%
\usepackage{mathrsfs}%
\usepackage[title]{appendix}%
\usepackage{xcolor}%
\usepackage{textcomp}%
\usepackage{manyfoot}%
\usepackage{booktabs}%
\usepackage{algorithm}%
\usepackage{algorithmicx}%
\usepackage{algpseudocode}%
\usepackage{listings}%
\usepackage{ulem}%
\usepackage{bm}

\theoremstyle{thmstyleone}%
\theoremstyle{thmstyletwo}%

\theoremstyle{thmstylethree}%

\let\polishl\l
\renewcommand{\l}{\lambda}

\newcommand{\be}{\begin{equation}}
\newcommand{\ee}{\end{equation}}
\newcommand{\bea}{\begin{eqnarray}}
\newcommand{\eea}{\end{eqnarray}}

\def\a{\alpha}

\def\G{\Gamma}
\def\d{\delta}

\def\e{\epsilon}

\def\l{\lambda}

\def\m{\mu}

\def\p{\pi}

\def\r{\rho}

\def\t{\tau}

\def\w{\omega}
\def\W{\Omega}

\def\blk{{\mathbf k}}

\def\blQ{{\mathbf Q}}

\def\callG{\mbox{$\mathcal{G}$}}

\def\1op{\hat{\mathbbm{1}}}
\def\nn{\nonumber}
\def\AA{\mathring{\mathrm{A}}}
\def\bz{{\mathbf z}}

\def\1op{\hat{\mathbbm{1}}}
\def\nn{\nonumber}

\def\bz{\mathbf 0}

\usepackage{newfloat}

\DeclareFloatingEnvironment[
    fileext=loe,
    listname={List of Extended Figures},
    name={Extended Data Fig.},
    placement=tbp
]{extendedfigure}

\begin{document}

\title[Article Title]{Quasiparticle phono-conversion: filming carriers coalescing into excitons}


\author[1,2]{\fnm{Enrico} \sur{Perfetto}}\email{eperfett@roma2.infn.it}

\author[3]{\fnm{Takumi} \sur{Fukuda}}

\author[3]{\fnm{Xing} \sur{Zhu}}

\author[3]{\fnm{Jacques} \sur{Hawecker}}

\author[3]{\fnm{Harley} \sur{Suchiang}}

\author[3,4]{\fnm{Joanna} \sur{Nadolna}}

\author[3]{\fnm{Nanami} \sur{Tomoda}}

\author[5]{\fnm{Suji} \sur{Park}}

\author[5]{\fnm{Houk} \sur{Jang}}

\author[6]{\fnm{Kenji} \sur{Watanabe}}

\author[7]{\fnm{Takashi} \sur{Taniguchi}}

\author[3]{\fnm{Michael K. L.} \sur{Man}}

\author[3]{\fnm{Julien} \sur{Mad\'eo}}

\author[3]{\fnm{Keshav M.} \sur{Dani}}\email{kmdani@oist.jp}

\author[1,2]{\fnm{Gianluca} \sur{Stefanucci}}\email{stefanuc@roma2.infn.it}

\affil[1]{Dipartimento di Fisica, Universit{\`a} di
Roma Tor Vergata, Via della Ricerca Scientifica 1,
00133 Rome, Italy}

\affil[2]{INFN, Sezione di Roma Tor Vergata, Via della Ricerca Scientifica
1, 00133 Rome, Italy}

\affil[3]{Femtosecond Spectroscopy Unit, Okinawa Institute of Science and Technology Graduate University, Onna, Okinawa, Japan 904-0495
}

\affil[4]{Laboratory of photocatalysis, Department of Environmental Technology, Faculty of Chemistry, University of Gdansk, 80–308 Gdansk, Poland
}

\affil[5]{Center for Functional Nanomaterials, Brookhaven National Laboratory, Upton, New York 11973, USA
}

\affil[6]{
Research Center for Electronic and Optical Materials, National Institute for Materials Science, 1-1 Namiki, Tsukuba 305-0044, Japan
}

\affil[7]{Research Center for Materials Nanoarchitectonics, National Institute for Materials Science, 1-1 Namiki, Tsukuba 305-0044, Japan
}


\abstract{
Condensed matter physics is replete with phenomena involving 
high-energy free particles coalescing into low-energy bound 
few-particle states. While the cooling of the 
individual particles is well understood, the crucial step by which 
cold free carriers form a bound state remains 
elusive, involving complex energy and 
momentum relaxation pathways.  
Here, by combining ultrafast time- and momentum-resolved 
photoemission spectroscopy  on a monolayer WSe$_2$ with 
the first-principles excitonic-Bloch equations, we 
resolve the conversion of initially free electrons and holes 
at the bandedges into bound excitons.    
With unprecedented energy resolution, we observe the transient 
\textit{coexistence} of free-carrier and excitonic bands, accompanied 
by a transfer of spectral weight between the two. 
We establish the phononic origin of exciton formation and ascribe 
this coexistence to a sequential relaxation cascade toward the 
lowest-energy excitonic states, wherein intermediate states remain 
weakly populated.     
The efficiency of this process  is controlled by valley 
multiplicity, large-momentum phonon emission and spin-flip 
processes.      
By elucidating how bound states emerge from their elementary 
constituents, our results point to strategies for engineering exciton 
formation, with direct implications for optical materials and devices 
that operate with excitons or free carriers. 

}


\maketitle

\section*{Introduction}
Exciton formation ~\cite{66rj-jbqw,PhysRevLett.81.2578,PhysRevLett.93.137401,trovatello2020,PhysRevB.42.7434,C6NR02516A,Steinleitner,D0CP03220D} lies at the core of light–matter interaction in 2D semiconductors, setting the fundamental limits for energy conversion and information processing~\cite{exrev1,RevModPhys.90.021001,exrev2,exrev3}. 
The efficiency and speed with which excitons emerge determine how effectively these materials can power technologies such as solar cells, photodetectors, and quantum optoelectronic devices. 

\begin{figure}[H]
\centering
\includegraphics[width=13cm]{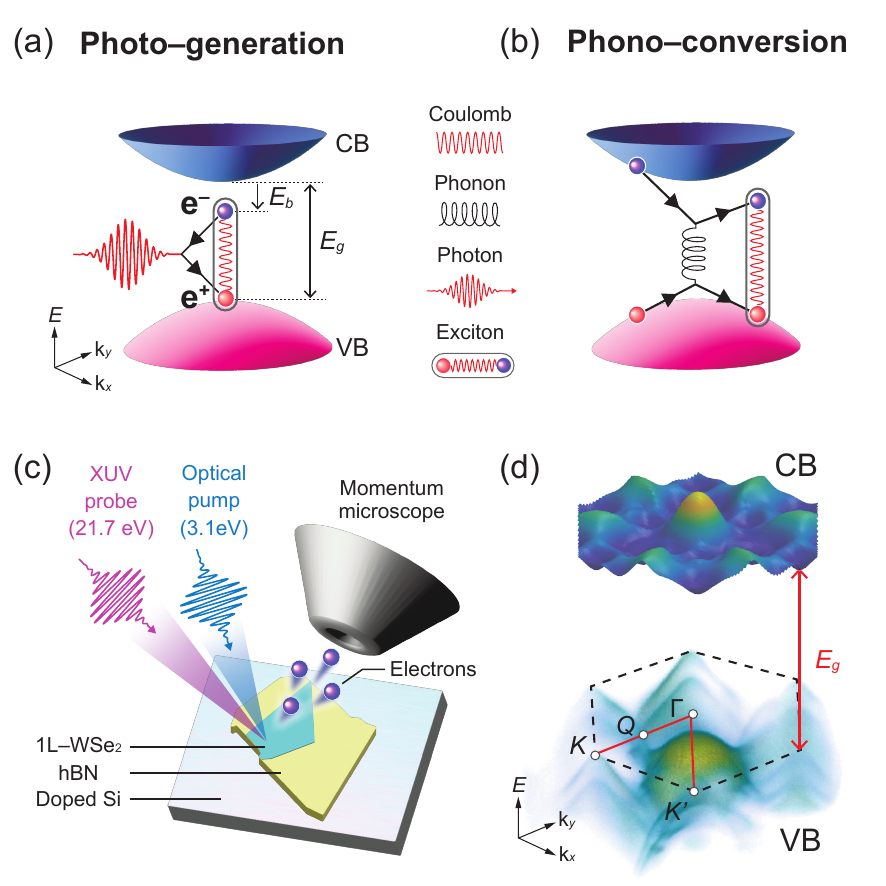}
\caption{{\bf Exciton phono-conversion dynamics and experimental setup.}
(a) Representation of the photo-generation process: an incoming photon (red pulse) with the same energy as the exciton is absorbed by an electron-hole pair to directly form a bound state. 
(b)  Representation of the phono-conversion process: an electron in the conduction band (blue sphere) and a hole in the valence band (red sphere) emit a phonon (spring) and lower their energy by forming a bound exciton. 
(c) Experimental setup of tr-ARPES for a monolayer (1L) of WSe$_2$ on a hBN layer put onto a doped Si substrate. The 3.1 eV and 21.7 eV are used as the optical pump and XUV probe, respectively, with a time delay. Photoelectrons are collected by the objective lens of the momentum microscope, enabling us to analyze the energy and momentum dynamics in the sample. 
(d)Three-dimensional view of the valence band (VB) and conduction band (CB) of 1L-WSe$_2$ in the first Brillouin zone. The VB was measured by equilibrium ARPES. The CB was computed from first-principles calculations. The bandgap $E_g = 2.06\ \mathrm{eV}$ corresponds to the energy difference between the VB maximum (VBM) and CB minimum (CBM) at the K valley of the first Brillouin zone. 
}
\label{fig1}
\end{figure}

Excitons can form via two primary pathways~\cite{66rj-jbqw}. 
The first is resonant (or direct) photoexcitation, in which photons with energies matching the exciton transition directly create excitons (Fig.~\ref{fig1}a). 
This conceptually simple mechanism, well-captured by mean-field theories~\cite{HAUG1985171,PhysRevB.37.941,PSMS.2019,PhysRevMaterials.5.083803}, has enabled the exploration of a wide array of emergent excitonic phenomena, including valley polarization~\cite{doi:10.1126/science.aac7820,polarization}, exciton-driven Floquet physics~\cite{doi:10.1073/pnas.2301957120,floquetnatphys}, exciton superfluidity~\cite{PhysRevLett.125.106401}, and bright-to-dark exciton scattering~\cite{doi:10.1126/science.aba1029,doi:10.1021/acs.nanolett.1c01839}. 

By contrast, the second mechanism, non-resonant (or indirect) 
excitation, is significantly more complex.    
Here, high-energy photons promote electrons and holes far above their 
respective band edges, generating hot quasi-free 
carriers which subsequently    
cool toward the band extrema through phonon emission—a process that 
is relatively well 
understood~\cite{caruso,perfetto_real-time_2023}.     
Exciton formation then occurs as electrons near the conduction band 
minimum  and holes near the valence band maximum 
 bind 
 through a cascade of phonon-mediated scatterings 
constrained by energy- and momentum-conservation 
(Fig.~\ref{fig1}b)~\cite{PhysRevB.65.035303}.    
Unraveling such a relaxation process is crucial for defining a 
fundamental timescale for hot carrier extraction before exciton 
formation in 2D material-based devices, thereby setting intrinsic 
efficiency limits for 2D-based optoelectronic 
technologies~\cite{Caruso_2026}.     

Despite its fundamental importance, the early-stage dynamics of exciton formation under non-resonant excitation remains largely inaccessible, and consequently far less understood.
Experimental observations so far have been restricted to time-resolved optical spectroscopy signals that capture the consequences of exciton formation~\cite{Steinleitner,C6NR02516A,cha,D0CP03220D,trovatello2020}, rather than the excitons themselves.
In fact, the population of excitons and free carriers cannot be 
directly disentangled and are instead inferred from phenomenological 
two-component models~\cite{PhysRevB.79.045320,condensation}.      
In principle, time- and angle-resolved photoemission spectroscopy (tr-ARPES) provides a direct route to track the conversion of free carriers into excitons with full momentum and energy resolution.
However, experimental limitations  have thus far hindered a direct and conclusive observation of this process~\cite{doi:10.1126/science.aba1029,gosetti2024unveiling}. 
Compounding the experimental challenges, the theoretical modeling of 
this nonequilibrium dynamics is equally formidable. 
Unlike resonant excitation, where excitons are generated directly, nonresonant excitation defies simple theoretical treatment capable to describe the multistep evolution of the transient mixture of unbound carriers and bound excitons. 
Accurately describing this relaxation  process requires demanding 
computational resources and the development of advanced many-body 
theoretical frameworks~\cite{10.21468/SciPostPhys.18.1.009}.

In this work, we directly observe the phonon-mediated coalescence of 
free electron–hole pairs into excitons in non-resonantly 
photoexcited monolayer (1L-) WSe$_2$.  
Using tr-ARPES experiments with unprecedented energy resolution 
(see Fig.~\ref{fig1}c for schematic setup and "Methods" for details), 
we resolve the transient \textit{coexistence} of free carriers and 
excitons  during the conversion process. 
Immediately following photoexcitation, the tr-ARPES spectrum reveals 
the characteristic low-energy shape of the conduction band populated 
by cooling carriers. 
Within a sub-picosecond timescale, the spectral weight associated 
with free carriers diminishes, while new, well-defined subbands at 
lower energies, attributable to excitonic states, emerge.   
To elucidate this process, we perform first-principles simulations 
based on the recently developed excitonic Bloch 
equations~\cite{10.21468/SciPostPhys.18.1.009}, which establish the 
phononic origin of exciton formation. The calculations show that the 
dynamics develops a pronounced bottleneck as carrier relaxation 
crosses over from a continuum of free-particle states to a discrete 
excitonic manifold. This crossover is highly nontrivial in 
two-dimensional transition-metal dichalcogenides because valley 
multiplicity gives rise to a dense manifold of optically bright and 
dark excitonic states of comparable energy~\cite{Deilmann_2019}. 
Populating these states requires large-momentum phonon emission and
spin-flip processes during carrier relaxation. 
Our findings indicate that intermediate excitonic states mediate the 
conversion from free carriers to bound excitons while maintaining a 
low population. This specific mechanism underlies the transient 
coexistence of free-carrier and low-energy excitonic states observed 
as distinct spectral signatures.

\section*{tr-ARPES of exciton phono-conversion}

The tr-ARPES measurement for 1L-WSe$_2$ was carried out using a time-of-flight momentum microscope (ToF-MM) at 100~K under ultrahigh vacuum conditions, where the schematic is drawn in Fig.~\ref{fig1}c, and the detail is described in the "Methods" part and Extended Data Fig.~\ref{EDexp1}. 
Free carriers are optically injected on the conduction band (CB) from the valence band (VB) using a 3.1-eV pump pulse, which is $\sim 1\ \mathrm{eV}$ above the energy gap $E_g$.
The pump fluence used was $1.6\ \mathrm{\mu J/cm^2}$, corresponding to a photoexcited carrier density of $3.2 \times 10^{11}\mathrm{cm^{-2}}$, which remains well below the Mott transition threshold to preserve excitonic stability~\cite{Chernikov2015,Steinhoff2017, Dendzik2020prl}. 
A time-delayed 21.7-eV probe pulse enables us to simultaneously capture both VB and CB as time-, energy-, and momentum-resolved photoemission spectra with $\sim 100~\mathrm{fs}$ temporal resolution.
Figure.~\ref{fig1}d shows the experimentally obtained three-dimensional band structure ($k_x$, $k_y$, and $E$) of VB before time zero and theoretically calculated CB.
The VB shows the negative parabolic dispersion at the $\Gamma$ point and $K(K')$ points hexagonally placed on the edge of the first Brillouin zone (see also Extended Data Fig.~\ref{EDexp1}c).
CB minima (valleys) are located at $K$ and $Q$ points.
The energy linewidth (FWHM) of the ARPES spectra was estimated to be $\sim 88\ \mathrm{meV}$ at the maximum of the VBs (see Extended Data Fig.~\ref{EDexp1}b and e).
This value is much less than the previously reported experiments for 1L-WSe$_2$ \cite{madeo2020, doi:10.1126/sciadv.abg0192}, allowing us to clearly distinguish spectral contributions between CB and excitons, whose energy gap is typically a few hundred meV.

\begin{figure}[H]
\centering
\includegraphics[width=12cm, clip]{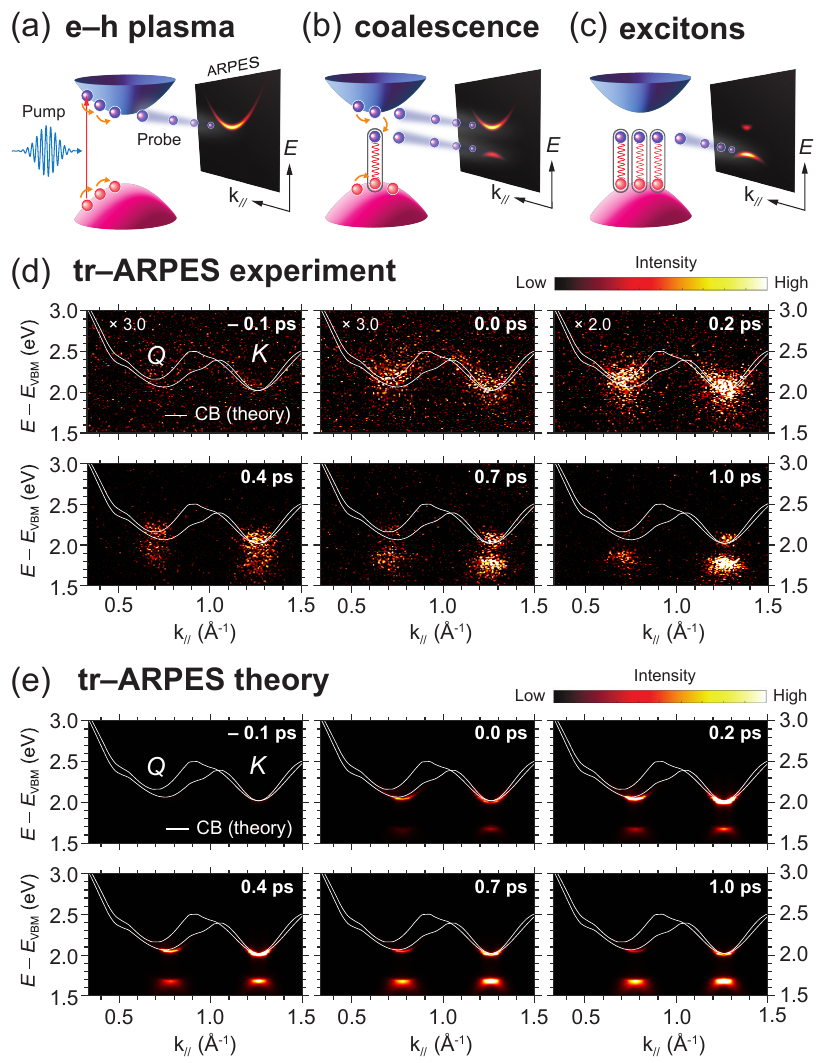}
\caption{{\bf Time-resolved ARPES in 1L-WSe$_{2}$.}
(a–c) Schematic view of exciton phono-conversion and its signatures in tr-ARPES: (a) Initial formation of an electron–hole plasma due to above-gap pumping and free-carrier relaxation toward the band edges, producing an ARPES signal that reflects the conduction band; (b) Coexistence of free carriers and bound excitons, giving rise to an additional excitonic sideband inside the gap; (c) Progressive coalescence of electron–hole pairs into excitons, resulting in a transfer of spectral weight from the conduction band signal to the in-gap excitonic sideband. 
(d) Experimental tr-ARPES data collected at different time delays from -0.1 to 1.0 ps  (e) Calculated tr-ARPES spectra at the corresponding time delays. A small energy broadening $\eta = 0.01\ \mathrm{eV}$ is used to resolve the dispersion of the excitonic sidebands. In all ARPES spectra, the energy $E$ is measured with respect to the VBM: $E - E_{\mathrm{VBM}}$. In the panels corresponding to the delay $= -0.1\ \mathrm{ps}$, the ab-initio spin-resolved CB structure is overlaid as a dashed white line.
}
\label{fig2}
\end{figure}

Non-resonant above-gap pump excitation generates a hot electron-hole (e-h) 
plasma. The e-h pairs initially redistribute within the respective 
band edges, then coalesce into bound excitons, and ultimately relax into the lowest energy exciton states (Fig.~\ref{fig2}a-c).
Here, we directly capture these electronic excitation and relaxation processes with tr-ARPES.
Figure~\ref{fig2}d displays a time-delay series of ARPES spectra around the CB energy level sliced along a $\Gamma \to K$ direction, including a $Q$ valley.
Although the excitation occurs within the $K$ valley, intervalley scattering mediated by the emission of large-momentum phonons efficiently transfers carriers to the $Q$ valleys already during the pumping process~\cite{https://doi.org/10.1002/ntls.10010,perfetto_real-time_2023},
see Fig.~\ref{fig2}d at delay $\tau = 0\ \mathrm{ps}$.
The positive parabolic dispersions at $K$ and $Q$ valleys are attributed to the presence of hot carriers, almost following the calculated CB dispersion.
Shortly after the pump pulse ends ($\tau = 0.2\ \mathrm{ps}$), free 
carriers accumulate near the band minima.
At later time delay, around $\tau = 0.4\ \mathrm{ps}$, transitions 
into the lowest-lying energy levels become apparent as free carriers begin to 
coalesce into bound states (see also Extended Data Fig.~\ref{EDfig1}).
Furthermore, the negative parabolic dispersion at 1.0 
ps also confirms that the lowest energy states arises from excitons 
(see Extended Data Fig.~\ref{EDexp3}) \cite{floquetnatphys, 
doi:10.1126/sciadv.abg0192}.  
The high-resolution experimental spectra reveal the  coexistence of 
free carriers and low-energy excitons, characterized by a progressive transfer 
of spectral weight from the conduction bands signal to the in-gap 
excitonic sidebands.    
This fundamental relaxation process terminates around  $\tau = 1.0\ \mathrm{ps}$.   

Real time
simulations successfully reproduce the experimental tr-ARPES spectra (see Methods), confirming the coexistence of  free carriers and excitons (Fig.~\ref{fig2}e).
As pointed out in Refs.~\cite{PhysRevB.94.245303,stefanucci2026unifiedfirstprinciplesformulatimeresolved,RustagiKemper2018,exph3}, the ARPES signal from  photoelectrons of momentum $\blk$ forming excitons occurs at energy 
$\epsilon_{v,\mathbf{k}-\mathbf{Q}} + E_{\mathbf{Q}}$, where $\epsilon_{v,\mathbf{k}}$ denotes the VB energy and $E_{\mathbf{Q}}$ the energy of excitons with center-of-mass momentum $\mathbf{Q}$, and has an intensity proportional to the population $N_{\blQ}$ of the corresponding exciton state. 
This implies that the selective population of bright excitons ($\mathbf{Q}=0$) results in the appearance of valence-band replicas blue-shifted by the exciton 
energy (see Fig.~\ref{fig2}b)~\cite{PhysRevB.94.245303,PSMS.2019,PBS.2020,RustagiKemper2018,PhysRevB.100.205401}, as recently demonstrated in photoexcited two-dimensional materials~\cite{doi:10.1126/sciadv.abg0192,https://doi.org/10.1002/ntls.10010}. 
The population of dark excitons with finite momentum $\blQ$ gives rise to replicas that are not only blue shifted but also displaced in momentum space.
The above-gap photoexcitation followed by relaxation populates a broad manifold of quasi-degenerate excitonic states~\cite{Deilmann_2019}.
Therefore, photoelectrons 
of same momentum $\mathbf{k}$ belong to excitons of different center-of-mass momentum (see Extended Data Fig.~\ref{EDfig2}).

The method used to simulate the real-time dynamics of free carrier and exciton populations is the 
first principles excitonic Bloch equations (XBE)~\cite{10.21468/SciPostPhys.18.1.009}, as described in Methods.   
In the XBE e-h states with center-of-mass momentum $\blQ$ are converted into exciton states with center-of-mass momentum $\blQ'$ by emitting a phonon with momentum $\blQ - \blQ'$.
Alternative pathways such as Coulomb-induced carrier-carrier scattering or 
Auger-like processes, which are known to become significant at higher carrier densities, are neglected. 
The agreement between theory and experiment in Fig.~\ref{fig2} unequivocally establishes phono-conversion as the dominant mechanism responsible for the formation of excitons.

The XBE  provide further microscopic insights into the phono-conversion process.
As electrons and holes approach the band edges, the 
dissipation of their excess energy shifts from transitions within a 
continuum of states to scattering into a discrete manifold of bound 
excitonic states. This transition severely restricts the available 
scattering phase space, resulting in a dynamical bottleneck during 
carrier relaxation, and ultimately manifesting as a marked carrier 
accumulation at the CB minima.
Energy conservation, combined with the relatively small phonon 
energies compared to the exciton binding energy, implies that 
exciton formation proceeds through a relaxation     
cascade~\cite{PhysRevB.65.035303,cascade,PhysRevResearch.4.043203}.
In this process, carriers sequentially relax along the descending 
energy landscape, ultimately populating the lowest-lying excitonic 
states at the high-symmetry 
points. Notably, the occupation of intermediate states remains low, 
which explains why higher-energy states (including the bright A 
exciton) are  not visible from the ARPES signal.    

We stress that the  existence of a regime in which free carriers and 
excitons coexist has remained uncertain, as neither 
experiments nor theoretical calculations had previously been able to 
establish it.     
Figure \ref{fig2} provides unambiguous evidence for 
such coexistence.

\section*{Phono-conversion timescale}

\begin{figure}[H]
\centering
\includegraphics[width=12cm, clip]{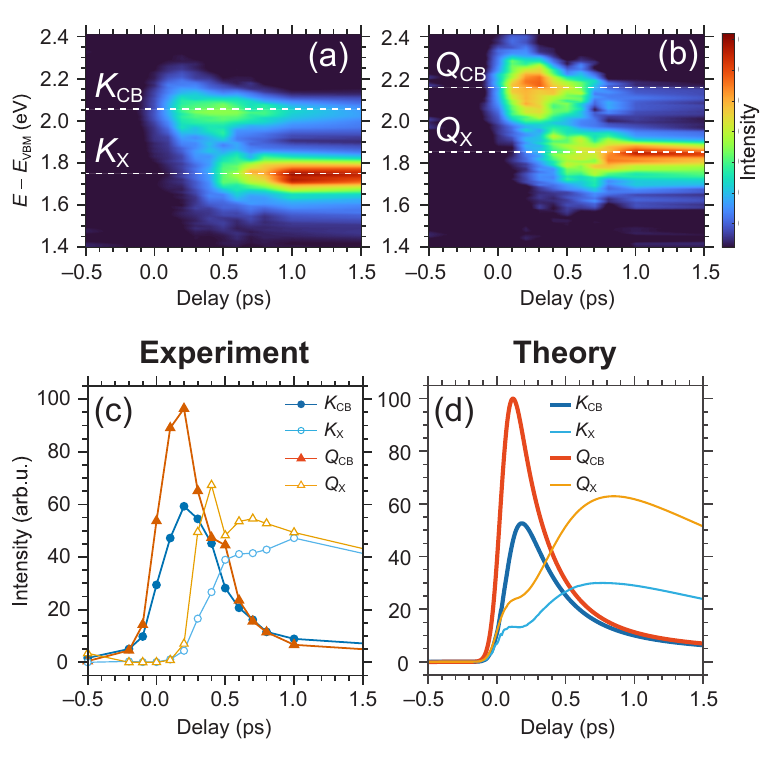}
\caption{{\bf Time evolution of transition from free carriers to excitons.}
The temporal energy evolution of photoelectron counts at the $K$ valley (a) and the $Q$ valley (b) extracted from the tr-ARPES measurement. 
The $K_{\mathrm{CB}}$ (2.06 eV) and $Q_{\mathrm{CB}}$ (2.16 eV) are the energy levels of CB at $K$ and $Q$ valleys, respectively.
The $K_{\mathrm{X}}$ (1.75 eV) and $Q_\mathrm{X}$ (1.85 eV) are the excitonic levels at $K$ and $Q$ valleys, respectively. (c) The dynamics of 
photoelectron intensity at the energy levels of $K_{\mathrm{CB}}$,  $Q_{\mathrm{CB}}$,  $K_{\mathrm{X}}$,  and $Q_\mathrm{X}$. extracted from 
(a) and (b). Connecting lines are guides to the eye. (d) Simulated tr-ARPES signal, processed identically to the experimental data in (c). 
Here, we have multiplied all curves by an exponential damping function with a lifetime of $\tau \approx 2\ \mathrm{ps}$ ~\cite{PhysRevB.93.205423}, 
accounting for the  electron-hole recombination that attenuates the quantities reported in panel (c).
}
\label{fig3}
\end{figure}

We analyze the time- and energy-resolved evolution of free carriers and 
excitons  at $K$ and  $Q$ valleys, as shown in 
Fig.~\ref{fig3}a-b, respectively   
(See details in Extended Data Fig.~\ref{EDexp2}).
Our results successfully distinguish two spectral peaks of CB and exciton energy levels at each time delay and valley.
The energy levels of CB and exciton at $K$ valley are at 
2.06 eV ($K_{\mathrm{CB}}$) and 1.75 eV ($K_{\mathrm{X}}$), 
while those in the $Q$ valley are at 2.16 eV ($Q_{\mathrm{CB}}$) and 1.85 eV ($Q_{\mathrm{X}}$).
Energy difference between $K$ and $Q$ valleys is $\sim 100\ \mathrm{meV}$.
In addition, the exciton binding energy is estimated to be $\sim 310\ \mathrm{meV}$.
These values are consistent with previous literature and our 
theoretical calculation~\cite{Deilmann_2019,wse2gw}, see Extended Data Fig.~\ref{EDfig2}a.
 As detailed in Extended Data Fig.~\ref{EDfig2}(b,c)
the signal at $K$ is predominantly attributed to electrons at
$K$ bound in spin-dark excitons with 
holes at $K$ and momentum-dark excitons with holes at
$K'$. Meanwhile, the signal at $Q$ stems mainly from electrons at Q 
bound in momentum-dark excitons with holes at $K$ and $K'$.
The formation of spin-dark excitons requires 
spin-flip processes and, according to our simulations,  
their growth is delayed and slower than that of spin-bright excitons, 
see Extended Data 
Fig.~\ref{SIfig1}. Nevertheless, the final populations of the two 
spin species is comparable, highlighting the crucial role of spin-flip 
scattering in shaping the final 
exciton distribution.

The measurements indicate that the colascence of e-h pairs into excitons initiates a few hundreds of femtoseconds after the pump excitation. 
In Fig. \ref{fig3}c, we evaluate the generation and relaxation dynamics of free carriers and excitons at $K$ and $Q$ valleys, extracted from Figs. \ref{fig3}a and b.
First, free carriers are generated in the  CBs and relax to the CB 
minima within $\sim 300\ \mathrm{fs}$.    
Subsequently, these carriers bound into excitons within a 
timescale of up to $\sim 1\ \mathrm{ps}$, consistent with optical-pump 
terahertz-probe   
experiments~\cite{Steinleitner,C6NR02516A,cha}. 
Finally, the excitons relax with a characteristic time constant of approximately 2~ps.

XBE simulations are reported  in Fig.~\ref{fig3} and well reproduce  the multiple characteristic timescales observed experimentally. 
The small discrepancy in the onset of exciton populations is due to a slightly lower photon energy used in the calculations (see Methods).

\section*{Dark excitons and lineshape asymmetry}

\begin{figure}[H]
\centering
\includegraphics[width=12cm, clip]{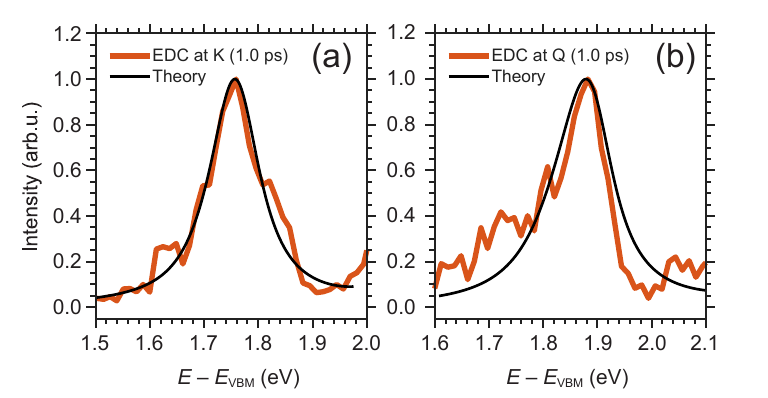}
\caption{
{\bf Asymmetry of lineshape at exciton peaks}.
Comparison of energy distribution curves (EDCs) around the exciton peaks between experimental and theoretical tr-ARPES data at 1.0 ps at the $K$ valley (a) and $Q$ valley (b).  
The experimental EDCs are extracted by integrating the ARPES signal within a momentum window of $0.037~\AA^{-1} \times 0.037~\AA^{-1}$ centered at the $K$ and $Q$ valleys.
Overlaid theoretical EDCs are obtained using the same integration procedure, computed with an energy broadening parameter $\eta=0.05$~eV. 
}
\label{fig4}
\end{figure}

The presence of multiple excitonic states leaves a 
distinct fingerprint in the photoemission lineshape, offering an 
additional benchmark for theory: a pronounced
asymmetry in the energy distribution curves (EDCs) at both $K$ and $Q$ 
points, characterized by an extended tail towards lower energies, see 
Fig.~\ref{fig4}a,b.

The asymmetry, which is absent under selective resonant 
excitation~\cite{floquetnatphys}, is accurately reproduced by our 
simulated tr-ARPES spectra.
The numerical analysis reveals that this feature is a general 
hallmark of a quasi-thermal exciton population, and it  originates 
from the imbalance between the effective masses $M_{\rm x}$ and $m_{v}$ of 
excitons and valence holes respectively    
-- $M_{\rm x}\approx 2m_{v}$ in WSe$_{2}$~\cite{Kormanyos_2015,PhysRevB.110.205417}.
Consider a photoelectron with momentum $\blk$ at the high-symmetry point, say the $K$ point. 
According to Extended Data Fig.~\ref{EDfig2}b, the tr-ARPES signal arises from 
 quasi-degenerate
excitonic states with momenta $\blQ$ around $\G$ and $K$, so that 
$E_{\blQ}\approx E_{\G/ K}+\frac{q^{2}}{2M_{\rm x}}$
(where $q$ is a small momentum). 
In both cases $\e_{v\blk-\blQ}\approx \e_{v K}-\frac{q^{2}}{2m_{v}}$, and therefore a signal
at energies  $\e_{v\blk-\blQ}+ E_{\blQ} \approx \e_{v K}+E_{\G/ K}+\frac{q^{2}}{2}(\frac{1}{M_{\rm x}}-\frac{1}{m_{v}}) \leq \e_{v K}+E_{\G/ K}$ with intensity proportional to the exciton population $N_{\blQ}$ 
is expected. Accordingly, 
the quasi-thermal distribution of excitons 
(Extended Data Fig.~\ref{EDfig1}c) gives rise to a lineshape featuring a peak near  $ \e_{v K}+E_{\G/ K}$, accompanied by a tail extending toward lower energies.
A similar analysis can be carried out for the lineshape at 
the $Q$ point, thus explaining the observed asymmetry in the EDCs (see also Extended Data Fig.~\ref{EDfig3}).

\section*{Summary and conclusions} 

Previous experimental~\cite{doi:10.1126/science.aba1029,gosetti2024unveiling} and theoretical~\cite{PhysRevB.94.245303,RustagiKemper2018,exph3} studies have clarified the spectral fingerprints of excitonic replicas in time-resolved ARPES. 
However, the dynamical coexistence of free carriers and excitons—and the microscopic timescale over which unbound e-h pairs coalesce into bound excitons—has remained inaccessible. Here, we provide the first direct, real-time observation of the phonon-mediated conversion of free electron–hole pairs into excitons, resolving the emergence of composite quasiparticles from their elementary constituents as it unfolds in time.  

Our results establish phonon-assisted electron–hole coalescence as a dominant and experimentally resolvable pathway for exciton formation in low-dimensional materials. First-principles simulations based on the excitonic Bloch equations framework reveal that exciton formation proceeds through multiple, well-separated timescales, reflecting a hierarchy of phonon-driven processes. These include intervalley scattering and spin-flip transitions, which govern the selective population of excitonic valleys at distinct symmetry points in momentum space.    
     
Altogether, our work establishes a powerful paradigm in which the combination of ultrafast spectroscopy and advanced theoretical modeling enables unprecedented insight into the role of bosonic interactions in driving emergent quantum phenomena in layered materials. 
Beyond advancing fundamental understanding, these findings offer critical and actionable knowledge for the rational design of next-generation optoelectronic and valleytronic devices that harness and control exciton dynamics at their most fundamental level.

\newpage

\section*{Methods} \label{sec11}

\bmhead{Sample preparation}
Monolayer(1L-) WSe$_2$ flakes were exfoliated from the bulk crystal (hqGraphene) onto a Si/SiO$_2$ substrate at the Quantum Material Press (QPress) facility in the Center for Functional Nanomaterials (CFN) at Brookhaven National Laboratory (BNL). Flakes of hexagonal boron nitride (hBN) were exfoliated from the bulk crystals (National Institute for Material Science, Japan) onto an $n$-doped silicon substrate. The monolayer flake was picked up from the Si/SiO$_2$ substrate and transferred on the top of the hBN flake using polycarbonate-based dry transfer technique, as shown in Extended Data Fig.~\ref{EDexp1}(a) and (b). 
The size of 1L-WSe$_2$ flake is estimated to be approximately $15\ \mathrm{\mu m} \times 20\ \mathrm{\mu m}$.
The sample was annealed in ultrahigh vacuum ($\sim 10^{-10}\ \mathrm{mbar}$) condition at $450\ ^{\circ}\mathrm{C}$ for 30 minutes.

\bmhead{Experimental set-up} 
The time-resolved ARPES experiment was carried out using a time-of-flight momentum microscope (ToF-MM, METIS1000, SPECS GmbH) combined with a home-built table-top high-harmonic generation (HHG) setup as an extreme ultraviolet (XUV) source, which has been described in detail in our previous work \cite{floquetnatphys}. 
We used a Yb-doped fiber laser operating at 2 MHz with 230-fs pulse width. 
The laser system delivered a pulse energy of 50 $\mathrm{\mu J}$ for 1030 nm, which was then split into 35 $\mathrm{\mu J}$ for XUV generation as an ARPES probe pulse and 15 $\mathrm{\mu J}$ for a non-collinear optical parametric amplifier (NOPA) to change the wavelength of the pump pulse. 
For the XUV generation, 35 $\mathrm{\mu J}$ (1030 nm) was frequency doubled by a $\mathrm{LiB_3O_5}$ (LBO) crystal to generate $\sim 15\ \mathrm{\mu J}$ (515 nm). The 515-nm light was then focused on a Kr gas jet to generate the 21.7-eV XUV probe. 
The XUV probe was finally collected and focused onto the sample inside the microscope using an ellipsoidal mirror. 
For the pump excitation, we used a NOPA system to generate a 400-nm (3.1-eV) pulse. 
The ToF-MM collected the photoemitted electrons from the sample using immersion objective lens under the XUV-probe irradiation conditions. 
A 10-$\mu$m field aperture was inserted at the image plane in the ToF-MM to collect photoelectrons from a clean monolayer region (Extended Data Fig. \ref{EDexp1}(b)).
The energy and momentum resolutions of the microscope were $< 30\ \mathrm{meV}$ and $< 0.01\ \AA^{-1}$, respectively. 
All the ARPES experiments were performed at 100 K under ultrahigh vacuum conditions ($\sim 10^{-10}\ \mathrm{mbar}$).

\bmhead{Data analysis} 
Our measurement provides three dimensional (3D) photoemission intensity data (Energy, $k_x$ and $k_y$). After correcting the distortion, we performed the analysis based on the corrected 3D data. All the photoemission spectra were normalized by the total photoemission signal of the valence bands of 1L-WSe$_2$ to compensate the variations in experimental integration time and photoemission yield. 
Extended Data Fig.~\ref{EDexp1}(c) shows the momentum image in valence band region ($E-E_{\mathrm{VBM}} = -0.40\ \mathrm{eV}$) before pump (-0.5 ps). $K$ points in a hexagonal pattern and $\Gamma$ point at the center can be seen. 
ARPES spectrum was obtained by cutting the 3D data along the $\Gamma$-$K$ direction, as shown in Extended Data Fig.~\ref{EDexp1}(d). The linewidth (FWHM) at the valence band maximum (VBM) is estimated to be 88 meV by the gaussian fitting (Extended Data Fig.~\ref{EDexp1}(e)).
To extract the population of free carriers in the conduction band and the excitons, the photoemission intensity weas integrated within a $0.23~\AA^{-1} \times 0.23~\AA^{-1}$ square region centered at $K$ valley and $Q$ valley to get EDC between 1.2 eV to 3.0 eV, as shown in Extended Data Fig.~\ref{EDexp2}.
Two distinct peaks attributed to the conduction band free carriers and excitons can be clearly resolved. The background (red line) is approximated by asymmetric least squares from the EDC before optical excitation (-0.5 ps data). After background subtraction, the spectra were fitted using a two-Gaussian model to separate the contribution from the conduction-band carriers (blue shaded region) and excitons (red shaded region). 
The overall fit (grey line) including the background shows good agreement with the EDC (blue markers). 
The populations of carriers and excitons were quantified from the area under the respective fitted curve.

\bmhead{First-principle simulations}

The nonequilibrium dynamics of carriers and excitons 
following laser excitation,
including relaxation mediated by phonon emission,
are simulated from first principles using the recently 
developed excitonic Bloch equations (XBE)~\cite{10.21468/SciPostPhys.18.1.009}.
These equations describe the temporal evolution of the 
occupations $N_{\l \blQ}$ of all  electron-hole (e-h) states in the system, 
including both quasi-free 
e-h pairs and bound excitons. Here $\blQ$ denotes the center-of-mass 
momentum of the e-h state, and $\l$ labels the corresponding branch.
The XBE formulation is based on two key concepts.
First, the  occupation $N_{\l \blQ}$ is decomposed into a coherent and an incoherent contribution
\be
N_{\l \blQ}=\d_{\blQ ,
{\bf{0}}}N^{{\rm coh}}_{\l}+N^{{\rm inc}}_{\l \blQ}
\ee
which obey distinct equations of motion.
The coherent component is given by the square-modulus of the
e-h polarization $N^{{\rm coh}}_{\l}=|\r_{\l}|^{2}$, and is nonzero only for
optically active bright states with zero momentum.
Second, the evaluation of both $\r_{\l}$ and $N^{{\rm inc}}_{\l \blQ}$
requires the explicit calculation of the occupations of {\it 
irreducible}  e-h states~\cite{PhysRevB.106.125403} 
$\widetilde{N}_{\l \blQ}$. These occupations enter the XBE 
as auxiliary dynamical variables and are essential to avoid overscreening effects,
ensuring a consistent treatment of many-body interactions.
The resulting XBE read~\cite{10.21468/SciPostPhys.18.1.009}
\bea
\frac{d}{dt}\r_{\l}&=&-iE_{\l \bz}\r_{\l}
-i\W_{\l}- \frac{1}{2}\sum_{\widetilde{\l}'\blQ}
\Big[\widetilde{\G}^{{\rm pol}}_{\l\widetilde{\l}'\blQ}
+\widetilde{\G}_{\l\widetilde{\l}'\blQ}
\widetilde{N}_{\widetilde{\l}'\blQ}
\Big]
\r_{\l} \nonumber \\
\frac{d}{dt}\widetilde{N}_{\widetilde{\l}\blQ}&=&
-\widetilde{\G}^{\rm out}_{\widetilde{\l}\blQ}\,
\widetilde{N}_{\widetilde{\l}\blQ}
+\widetilde{\G}^{\rm in}_{\widetilde{\l}\blQ}\,
\big(1+\widetilde{N}_{\widetilde{\l}\blQ}\big) \nonumber \\
\frac{d}{dt}N^{\rm inc}_{\l\blQ}&=&
-\G^{\rm out}_{\l\blQ}\,
N^{\rm inc}_{\l\blQ}
+\G^{\rm in}_{\l\blQ}\,
\big(1+N^{\rm inc}_{\l\blQ}\big)
+\sum_{\l'\widetilde{\l}}|S^{\blQ}_{\l\widetilde{\l}}|^{2}
\Big[\widetilde{\G}^{\rm pol}_{\l'\widetilde{\l}\blQ}
+\widetilde{\G}_{\l'\widetilde{\l}\blQ}\widetilde{N}_{\widetilde{\l}\blQ}
\Big]|\r_{\l'}|^{2}
\label{xbe}
\eea
In the above equations $E_{\l \blQ}$ is the energy of the 
e-h states, obtained from the solution of the Bethe-Salpeter
equation (BSE)
\begin{align}
(\e_{c\blk+\blQ}-\e_{v\blk})A^{\l\blQ}_{cv\blk}-
\sum_{c'v'\blk'}
K^{\rm HSEX,\blQ}_{cv\blk,c'v'\blk'}A^{\l\blQ}_{c'v'\blk'}=
E_{\l\blQ}A^{\l\blQ}_{cv\blk},
\label{eigeneqX}
\end{align}
with $\e_{v(c)\blk}$ the valence (conduction) band dispersion, 
$K^{\rm HSEX}$ the Hartree plus screened-exchange (HSEX) kernel, 
and $A^{\l\blQ}$ the e-h wavefunction.
The discrete sector of the BSE spectrum describes bound 
excitonic states, while the continuum 
corresponds to unbound, quasi–free e-h pairs.
The quantity $S^{\blQ}$ denotes the overlap between the 
wavefunctions  $A^{\l\blQ}$ obtained from Eq.~(\ref{eigeneqX}) and the 
wavefunctions of irreducible p-h states. The latter are computed by 
solving an analogous BSE equation, 
in which the HSEX kernel is replaced by the screened-exchange 
one~\cite{PhysRevB.106.125403}.
The laser-induced driving force $\W_{\l}=\sum_{cv\blk} A^{\l\bf{0}}_{cv\blk} \W_{cv\blk} $
is determined by the Rabi frequencies $ \W_{cv\blk}$ corresponding to 
vertical transitions from  valence states $v\blk$ 
to the conduction states $c\blk$.
The XBE include also several  rates   describing
distinct phonon-mediated processes.
These comprise the 
polarization-decay 
($\widetilde{\G}^{{\rm pol}}$), and inelastic phonon-assisted 
transitions between e-h states.
The latter accounts for both the depletion of a given e-h state, 
through out-scattering processes ($\G^{{\rm 
out}}$, $\widetilde{\G}^{{\rm out}}$) and its increasing population via 
in-scattering processes
($\G^{{\rm in}}$, $\widetilde{\G}^{{\rm in}}$, $\widetilde{\G}$). 
These rates are determined by the overlap of electron–hole wavefunctions,
modulated by the corresponding electron–phonon matrix elements,
and are constrained by energy 
conservation during the scattering process. Explicit expressions 
for all rates are provided in the Supplementary Information.

Based on the occupations $N_{\l \blQ}(\t)$ given by the XBE,
the time-resolved ARPES signal 
associated with photoelectrons of energy $\e$ emitted by a probe 
pulse with central photon energy $\w_{0}$ at pump-probe delay $\t$
can be expressed as~\cite{stefanucci2026unifiedfirstprinciplesformulatimeresolved}
\begin{align}
I_{\blk}(\t,\e)&\propto \sum_{\l\blQ}N_{\l\blQ}(\t)\sum_{v}
\Big|\sum_{c}A^{\l\blQ}_{cv\blk-\blQ}\Big|^{2}
\nn \d(\e_{v\blk-\blQ}+E_{\l\blQ}-\e+\w_{0}),
\label{tdphincoh1}
\end{align}
where, for simplicity, photoemission matrix elements 
have been neglected.

The single-particle band structure, phonon dispersions, electron–phonon couplings, 
and equilibrium excitonic properties of monolayer  WSe$_{2}$
were computed following the  methodology  of 
Refs.~\cite{perfetto_real-time_2023,stefanucci2026unifiedfirstprinciplesformulatimeresolved}.
Within this framework, the energies and wavefunctions of irreducible
e-h states coincide with those obtained from Eq.~(\ref{eigeneqX}).
Consequently, the exciton–phonon and irreducible-exciton–phonon couplings 
are identical.
All quantities were evaluated and stored on a $\blk$-grid of 3072 points 
 sampling the first Brillouin zone.
The electronic subspace was truncated to the two highest valence 
bands and the two lowest conduction bands, 
while the excitonic subspace was restricted to the ten lowest bound 
excitonic branches obtained from Eq.~(\ref{eigeneqX}).
To reduce memory requirements, the quasi-free e-h
continuum eigenstates of Eq.~(\ref{eigeneqX}) were approximated as
 $E_{\l\blQ} \approx \e_{c_{\l}\blk+\blQ}-\e_{v_{\l} 
\blk_{\l}}$, and 
$A^{\l\blQ}_{cv\blk} \approx 
\d_{c,c_{\l}}\d_{v,v_{\l}}\d_{\blk,\blk_{\l}} $.
In the simulations, we explicitly included the lowest 200 e-h
bands in the continuum sector, spanning energies up to
$2.7$~eV (approximately $\sim 0.7~$eV above than the bandgap).
Exciton–phonon couplings were stored on a coarser $\blk$-grid of 192 points
and interpolated on-the-fly onto the denser grid during the time-dependent simulations.

We initialize the system in thermal equilibrium at $T=100~$K,
by setting $\r_{\l}=N_{\l \blQ}=\widetilde{N}_{\l \blQ}=0$.
The dynamics are triggered by illuminating  the system with a $\sim 
250~$fs pump pulse lineraly polarized along the $x$ direction, with 
above-gap central frequency $\sim 2.4$~eV,
and of sufficiently low intensity to generate an excitation density $\sim 
10^{11}~{\rm cm}^{-2}$.
The XBE are integrated using a fourth-order Runge-Kutta solver 
with time-step of $0.1~$fs.


\backmatter

\section*{Supplementary information}

\subsection{Scattering rates}

In this Section, we detail the evaluation of the scattering
rates entering the XBE introduced in the Methods.
The relevant ingredients are  the phonon frequencies $\w_{\a \blQ}$, phonon 
occupation numbers $n_{\a\blQ}$, electron-phonon (e-p) 
matrix elements, and the energies and wavefunctions of electron–hole 
(e–h) states.
We adopt the convention that the e–p coupling $g^{\m \m'}_{\a -\blQ}(\blk)$ 
describes the amplitude for an electron 
with momentum $\blk$ to be scattered from band $\m$ to band
$\m'$ by the phonon mode $\a$ with momentum $\blQ$.
Electron–hole eigenmodes are obtained by solving the Bethe–Salpeter equation (BSE)
\be
 (\e_{c\blk+\blQ}-\e_{v\blk})A^{\l\blQ}_{cv\blk}-
 \sum_{c'v'\blk'}
 K^{\rm HSEX,\blQ}_{cv\blk,c'v'\blk'}A^{\l\blQ}_{c'v'\blk'}=
 E_{\l\blQ}A^{\l\blQ}_{cv\blk},
 \label{bse}
 \ee
where $\e_{v(c)\blk}$ denotes the valence (conduction) band 
dispersion, and $A^{\l\blQ}$ is the e-h wavefunction
associated with eigen-energy $E_{\l\blQ}$.
The interaction kernel
$K^{\rm HSEX,\blQ}_{cv\blk,c'v'\blk'}=
W_{c\blk+\blQ\,v'\blk'\,v\blk\,c'\blk'+\blQ}-
v_{c\blk+\blQ\,v'\blk'\,c'\blk'+\blQ\,v\blk}$ 
includes both the Hartree and screened-exchange contributions.
Here, $v$ denotes the bare Coulomb
interaction, while $W$ is the statically screened Coulomb interaction.
Irreducible e–h states are instead obtained from a modified BSE
 \be
 (\e_{c\blk+\blQ}-\e_{v\blk})\widetilde{A}^{\l\blQ}_{cv\blk}-
 \sum_{c'v'\blk'}
 K^{\rm SEX,\blQ}_{cv\blk,c'v'\blk'}\widetilde{A}^{\l\blQ}_{c'v'\blk'}=
 \widetilde{E}_{\l\blQ}\widetilde{A}^{\l\blQ}_{cv\blk},
 \label{irrbse}
 \ee
where the kernel retains only the screened-exchange contribution, 
i.e. $K^{\rm 
SEX,\blQ}_{cv\blk,c'v'\blk'}=W_{c\blk+\blQ\,v'\blk'\,v\blk\,c'\blk'+\blQ}$.
The quantities introduced above allow us to construct
the exciton–phonon coupling matrix elements~\cite{exph3,PhysRevB.105.085111}
\bea
\callG^{\l\l'}_{\a-\blQ'}(\blQ)&=& 
\sum_{c_{1}c_{2}v_{1}\blk_{1}}
A^{\l\blQ\ast}_{c_{1}v_{1}\blk_{1}}g^{c_{1}c_{2}}_{\a-\blQ'}(\blk_{1}+\blQ)
A^{\l'\blQ-\blQ'}_{c_{2}v_{1}\blk_{1}} \nonumber \\
&-&
\sum_{c_{1}v_{1}v_{2}\blk_{1}}A^{\l\blQ\ast}_{c_{1}v_{1}\blk_{1}}
g^{v_{2}v_{1}}_{\a-\blQ'}(\blk_{1}+\blQ')A^{\l'\blQ-\blQ'}_{c_{1}v_{2}\blk_{1}+\blQ'}
\label{callGx2}
\eea
as well as the irreducible-exciton–phonon couplings~\cite{10.21468/SciPostPhys.18.1.009}
\bea
 \widetilde{\callG}^{\widetilde{\l}\l'}_{\a-\blQ'}(\blQ)&=& 
 \sum_{c_{1}c_{2}v_{1}\blk_{1}}
 \widetilde{A}^{\widetilde{\l}\blQ\ast}_{c_{1}v_{1}\blk_{1}}g^{c_{1}c_{2}}_{\a-\blQ'}(\blk_{1}+\blQ)
 A^{\l'\blQ-\blQ'}_{c_{2}v_{1}\blk_{1}} \nonumber \\
&-&
 \sum_{c_{1}v_{1}v_{2}\blk_{1}}\widetilde{A}^{\widetilde{\l}\blQ\ast}_{c_{1}v_{1}\blk_{1}}
 g^{v_{2}v_{1}}_{\a-\blQ'}(\blk_{1}+\blQ')A^{\l'\blQ-\blQ'}_{c_{1}v_{2}\blk_{1}+\blQ'}.
 \label{callGx2irrQ}
 \eea
These effective couplings $\callG$ ($\widetilde{\callG}$) describe the amplitude for an e-h state 
in branch $\l$ and center-of-mass momentum $\blQ$ to scatter into
a (irreducible) state in branch $\l'$ and momentum $\blQ-\blQ'$ by exchanging
a phonon in mode $\a$.
Once the exciton–phonon couplings are known, the rates are evaluated according 
to~\cite{10.21468/SciPostPhys.18.1.009}
\begin{align}
\widetilde{\G}^{\rm pol}_{\l\widetilde{\l}'\blQ}&=
2\p\sum_{\a}\frac{|\widetilde{\callG}^{\widetilde{\l}'\l}_{\a -\blQ}(\blQ)|^{2}}{2\w_{\a\blQ}}
\Big[\d(\widetilde{E}_{\widetilde{\l}'\blQ}-E_{\l\bz}+\w_{\a\blQ})(1+n_{\a-\blQ})+
\d(\widetilde{E}_{\widetilde{\l}'\blQ}-E_{\l\bz}-\w_{\a\blQ})n_{\a\blQ}\Big]. 
\nonumber \\ 
\widetilde{\G}_{\l\widetilde{\l}'\blQ}&=
2\p\sum_{\a}\frac{|\widetilde{\callG}^{\widetilde{\l}'\l}_{\a -\blQ}(\blQ)|^{2}}{2\w_{\a\blQ}}
\Big[\d(\widetilde{E}_{\widetilde{\l}'\blQ}-E_{\l\bz}+\w_{\a\blQ})-
\d(\widetilde{E}_{\widetilde{\l}'\blQ}-E_{\l\bz}-\w_{\a\blQ})\Big] 
\nonumber \\ 
\widetilde{\G}^{\rm out}_{\widetilde{\l}\blQ}&=2\p
\sum_{\l'\a\blQ'}
\frac{\big|\widetilde{\callG}^{\widetilde{\l}\l'}_{\a-\blQ'}(\blQ)\big|^{2}}{2\w_{\a\blQ'}}
\big(1+N^{\rm inc}_{\l'\blQ-\blQ'}\big)
\nn\\
&\times\Big[
\d\big(E_{\l'\blQ-\blQ'}-\widetilde{E}_{\widetilde{\l}\blQ}+\w_{\a\blQ'}\big)(1+n_{\a\blQ'})
+\d\big(E_{\l'\blQ-\blQ'}-\widetilde{E}_{\widetilde{\l}\blQ}-\w_{\a\blQ'}\big)n_{\a-\blQ'}\Big],
\nonumber \\ 
\widetilde{\G}^{\rm in}_{\widetilde{\l}\blQ}&=2\p
\sum_{\l'}\sum_{\blQ'\a}
\frac{\big|\widetilde{\callG}^{\widetilde{\l}\l'}_{\a-\blQ'}(\blQ)\big|^{2}}{2\w_{\a\blQ'}}
N^{\rm inc}_{\l'\blQ-\blQ'}
\nn\\
&\times \Big[
\d\big(E_{\l'\blQ-\blQ'}-\widetilde{E}_{\widetilde{\l}\blQ}+\w_{\a\blQ'}\big)n_{\a\blQ'}
+\d\big(E_{\l'\blQ-\blQ'}-\widetilde{E}_{\widetilde{\l}\blQ}-\w_{\a\blQ'}\big)(1+n_{\a-\blQ'})\Big].
\end{align}
The rates $\G^{{\rm in/out}}$ have the same 
functional form as the corresponding irreducible rates
$\widetilde{\Gamma}^{\mathrm{in/out}}$, with the irreducible 
exciton energies $\widetilde{E}$
and couplings $\widetilde{\mathcal{G}}$ replaced by the
exciton energies $E$ and couplings $\mathcal{G}$, respectively~\cite{10.21468/SciPostPhys.18.1.009}. 
Both $\Gamma^{\mathrm{in/out}}$ and 
$\widetilde{\Gamma}^{\mathrm{in/out}}$ depend on the time-dependent 
state occupations $N^{\rm inc}_{\l'\blQ}$ and must therefore be re-evaluated at each time step.

\section*{Data availability}

The data supporting the findings of this study are available upon request.

\section*{Acknowledgements}
E.P. and G.S. acknowledge funding from Ministero Università e Ricerca PRIN under grant agreement No. 2022WZ8LME, from INFN through project TIME2QUEST and 
from the Horizon Europe research and innovation program of
the European Union under the Marie Sk\polishl{}odowska-Curie
grant agreement 101118915 (TIMES).
T.F. acknowledges support from the JSPS KAKENHI (Grant No. 25K17332) and Sasakawa Scientific Research Grant from the Japan Science Society (Grant No. 2024-2030).
J.M. acknowledges support from the JSPS KAKENHI (Grant No. 24K00561).
K.M.D acknowledges support from the JSPS KAKENHI (Grant Nos. 23K25807 and 24H00191) and JST FOREST program (Grant No. JPMJFR2230).
J.N. acknowledges support from the NAWA Bekker Programme (Project No. BPN/BEK/2023/1/00009/DEC/1).
This research used the Quantum material press (QPress) of the Center for Functional Nanomaterials (CFN), which is a U.S. Department of Energy Office of Science User Facility, at Brookhaven National Laboratory under Contract No. DE-SC0012704.
K.W. and T.T. acknowledge support from the JSPS KAKENHI (Grant Numbers 21H05233 and 23H02052), the CREST (Grant No. JPMJCR24A5), JST and World Premier International Research Center Initiative (WPI), MEXT, Japan.

\section*{Author contribution}

T.F. and X.Z. performed the TR-ARPES measurements with the assistance 
of N.T., H.S., M.K.L.M. and J.M.,
and under the supervision of K.M.D.
T.F., X.Z. and J.H. performed the analysis of the data under the supervision of K.M.D.
E.P. and G.S.
implemented the theoretical simulations.
H.S., J.N., S.P., H.J., K.W. and T.T prepared the 
samples.
E.P., G.S. and K.M.D. conceived and designed the study. All authors contributed 
to the writing and reviewing of the paper.

\section*{Competing interests}

J.M., M.K.L.M. and K.M.D. are inventors on a granted patent related to this work filed by the Okinawa Institute of Science and Technology School Corporation (US patent 11,372,199). The authors declare no other competing interests.

\newpage

\begin{extendedfigure}[H]
\centering
\includegraphics[width=13cm, clip]{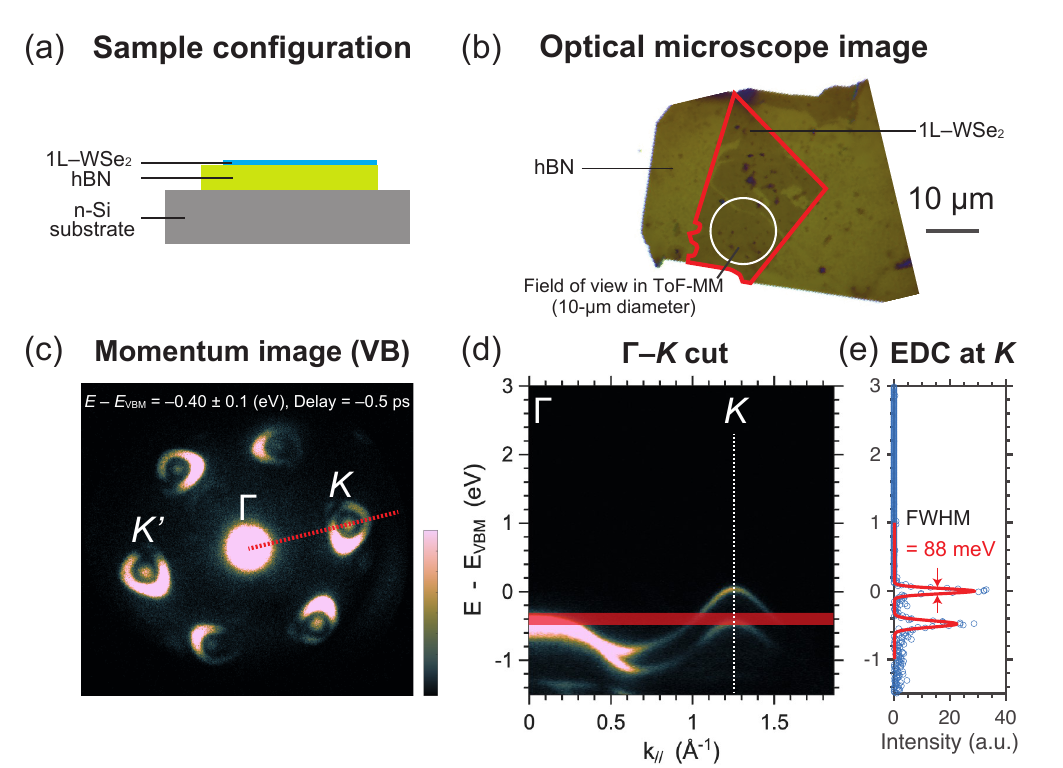}
\caption{{ \bf $|$ Sample configurations and static tr-ARPES data of 1L–WSe$_2$}
(a) Sample configuration of 1L–WSe$_2$ on hBN. 
(b) Optical microscope image of the sample. The brown area is hBN. The red marked area is 1L-WSe$_2$ on the hBN. The white circle denotes the field of view with 10-$\mu$m diameter during tr-ARPES measurements using the ToF-MM.    
(c) Momentum image of 1L-WSe$_2$ around valence band (VB) maximum at a negative time delay (-0.5 ps). The image was cut at $E - E_{\mathrm{VBM}} = -0.40\ \mathrm{eV}$ integrated by $\pm 0.1\ \mathrm{eV}$, which is shown in the red shaded area in (d). $K$ points are located in a hexagonal pattern. $\Gamma$ point is at center.  
(d) Static ARPES spectrum along $\Gamma$-$K$ direction (red dashed line in (c)). 
(e) Energy distribution curve (EDC) at $K$ point with the VB fit using gaussian functions. The EDC is extracted by integrating $0.022~\AA^{-1} \times 0.022~\AA^{-1}$ square region in momentum space centered at the $K$ point. Linewidth (FWHM) of the top of VB was estimated to be 88 meV.
}
\label{EDexp1}
\end{extendedfigure}

\begin{extendedfigure}[H]
\centering
\includegraphics[width=14cm, clip]{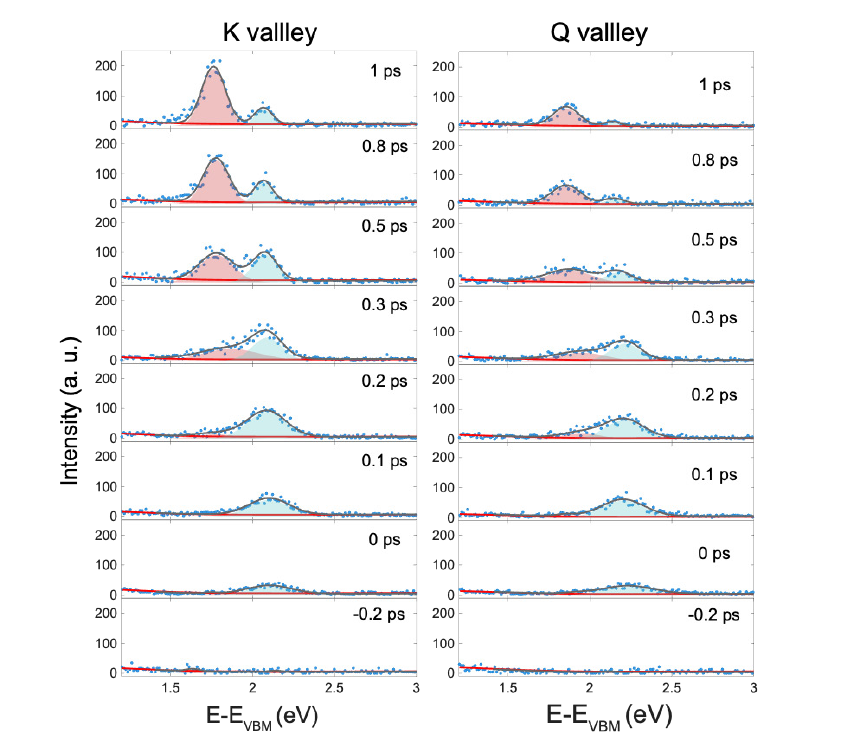}
\caption{{ \bf $|$ Extracting population dynamics of conduction bands and excitons}
To extract the population dynamics, EDCs for each delay was extracted by integrating $0.23~\AA^{-1} \times 0.23~\AA^{-1}$ square region centered at $K$ and $Q$ valleys in momentum space. Two gaussian fitting functions nicely caught two distinct peaks od conduction bands and excitons for each valley. Areas of the photoemission counts are plotted as population dynamics in Fig.~\ref{fig3}(c). 
}
\label{EDexp2}
\end{extendedfigure}

\begin{extendedfigure}[H]
\centering
\includegraphics[width=8cm, clip]{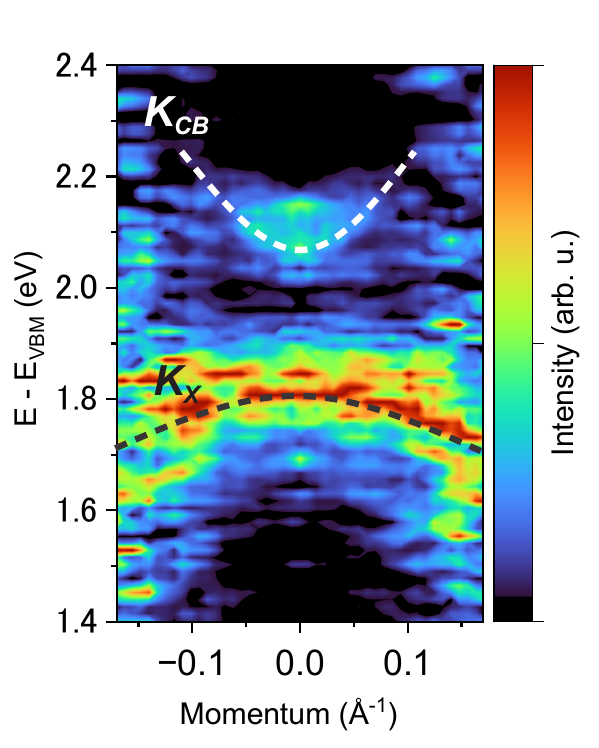}
\caption{{ \bf $|$ Dispersion of the conduction band and exciton at $K$ valley at 1.0 ps}
Two distinct peaks, as shown in Extended Data Fig.\ref{EDexp2}, have 
energy dispersions in momentum space. White and black dashed lines are guide for the eyes. The conduction band $K_{\mathrm{CB}}$ shows positive dispersion, which is more prominent at earlier time delays (see Fig.~\ref{EDexp2}(d)). The exciton $K_{\mathrm{X}}$ shows negative dispersion, which is obviously seen at the later time delay when the spectrum comes to have a minimum broadening effect. 
}
\label{EDexp3}
\end{extendedfigure}

\begin{extendedfigure}[H]
\centering
\includegraphics[width=6cm, clip]{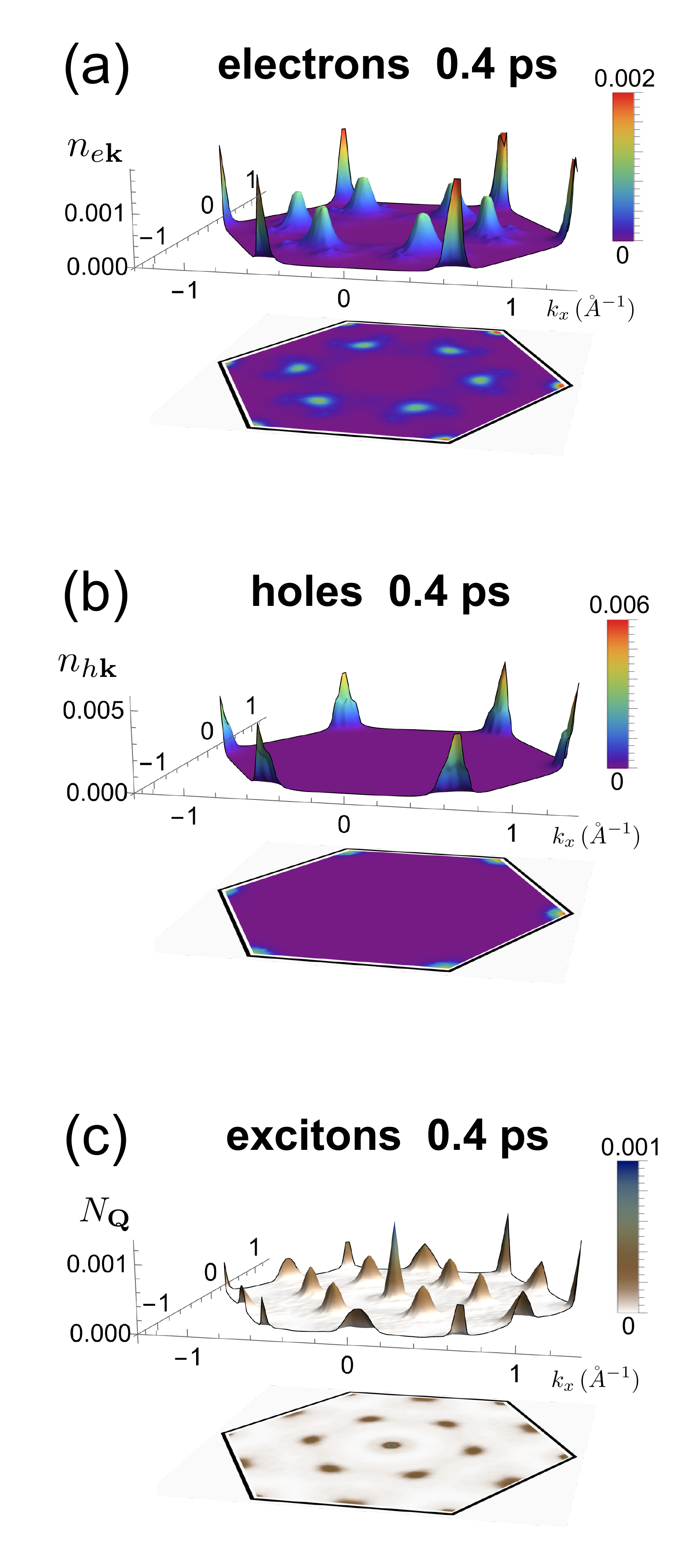}
\caption{{ \bf $|$ Momentum-resolved electronic and 
excitonic populations at coalescence.}
(a) Electron distribution $n_{e\blk}$, 
 (b) hole distribution
 $n_{h\blk}$, and 
 (c) exciton distribution $N_{\blQ}$
 calculated across the entire first Brillouin zone at time delay 
 $\tau=0.4$~ps. 
At this stage, the e-h plasma has largely dissipated 
its excess kinetic energy. Further relaxation
of the electronic subsystem occurs predominantly through exciton formation.
Panel~(c) shows 
a quasi-thermal distribution of excitons
around the high-symmetry points 
$\Gamma,\;K,\;Q,\;M$. Since all these excitons are 
formed with valence holes located in either the $K$ or  $K'$ 
valleys, the prominent ARPES signal comes from
photoelectron with momenta $\blk$
near the $K$ and $Q$ points, see Extended Data Fig.~\ref{EDfig2}.
The spectra in Fig.~\ref{fig2}d-e are consistent with this analysis.                    
}
\label{EDfig1}
\end{extendedfigure}

\newpage

\begin{extendedfigure}[H]
\centering
\includegraphics[width=10cm, clip]{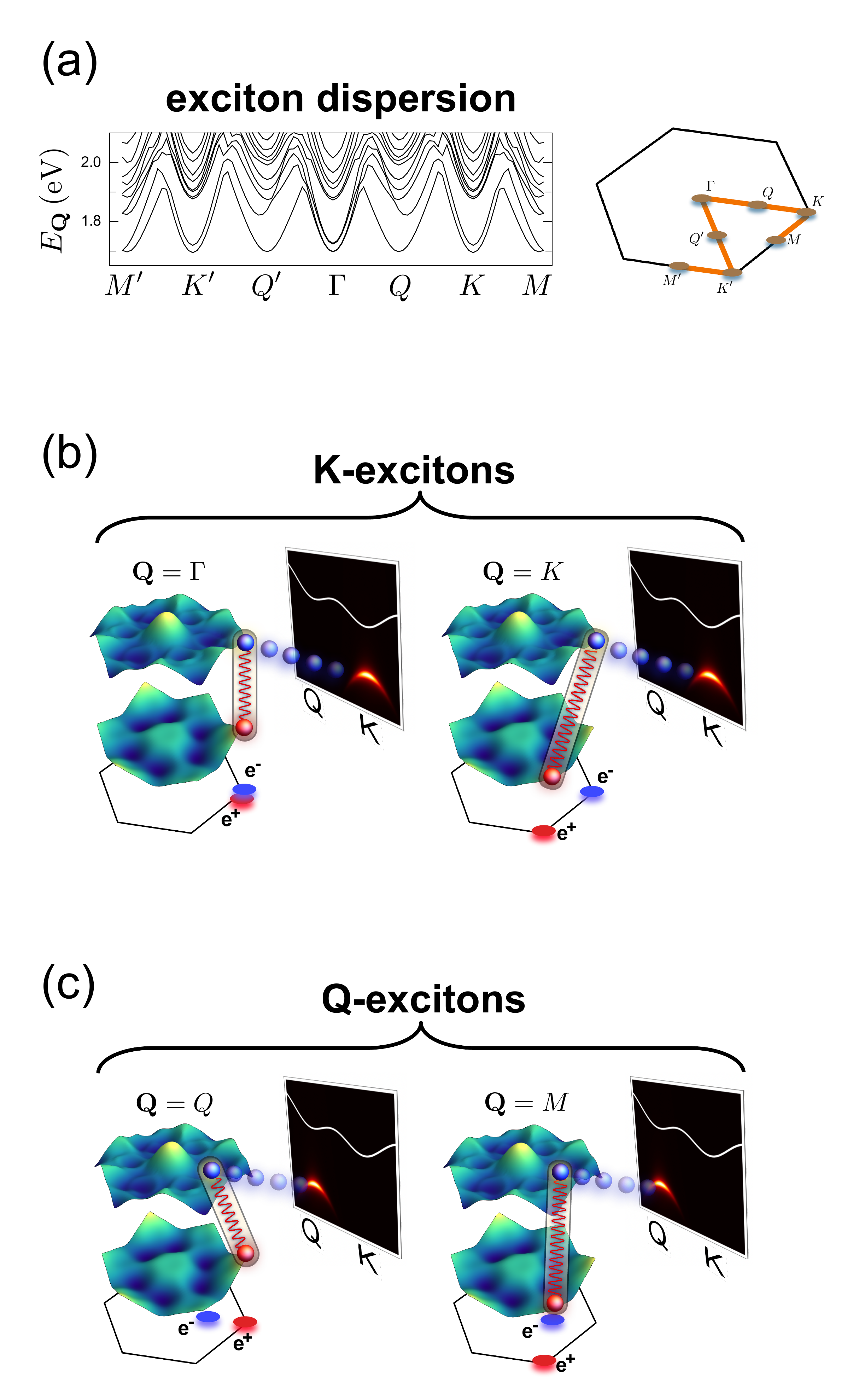}
\caption{{\bf $|$ Disentangling dark excitons 
contributions to ARPES signal.}
(a) Calculated low-energy exciton dispersion along a high-symmetry momentum path in the
first Brillouin zone, as indicated in the inset. 
(b)
Schematic illustration of the microscopic structure of the 
$K$-excitons, i.e. the excitonic states  contributing to the ARPES
signal near $\blk= K$.
These include direct excitons with center-of-mass momentum  $\blQ \approx \Gamma$
where both the electron and hole reside at the $K$ point, and indirect excitons with 
$\blQ \approx K$
composed of a hole at $K'$ and an electron at $K$;
(c) Same as (b), for the  
$Q$-excitons, which contribute to the ARPES signal near
$\blk= Q$.
These are indirect excitons with $\blQ \approx 
Q(M)$ comprising a hole at $K$ ($K'$)  and an 
electron at $Q$.
The signal observed at the $K$ valley
arises from a manifold of excitons having $\blQ$ near the $\Gamma$ and
$K$ points. 
Similarly, the signal observed at the $Q$ valley 
stems from  indirect excitons with $\blQ$ near the $Q$ and  
$M$ points.                    
}
\label{EDfig2}
\end{extendedfigure}

\newpage

\begin{extendedfigure}[H]
\centering
\includegraphics[width=8cm, clip]{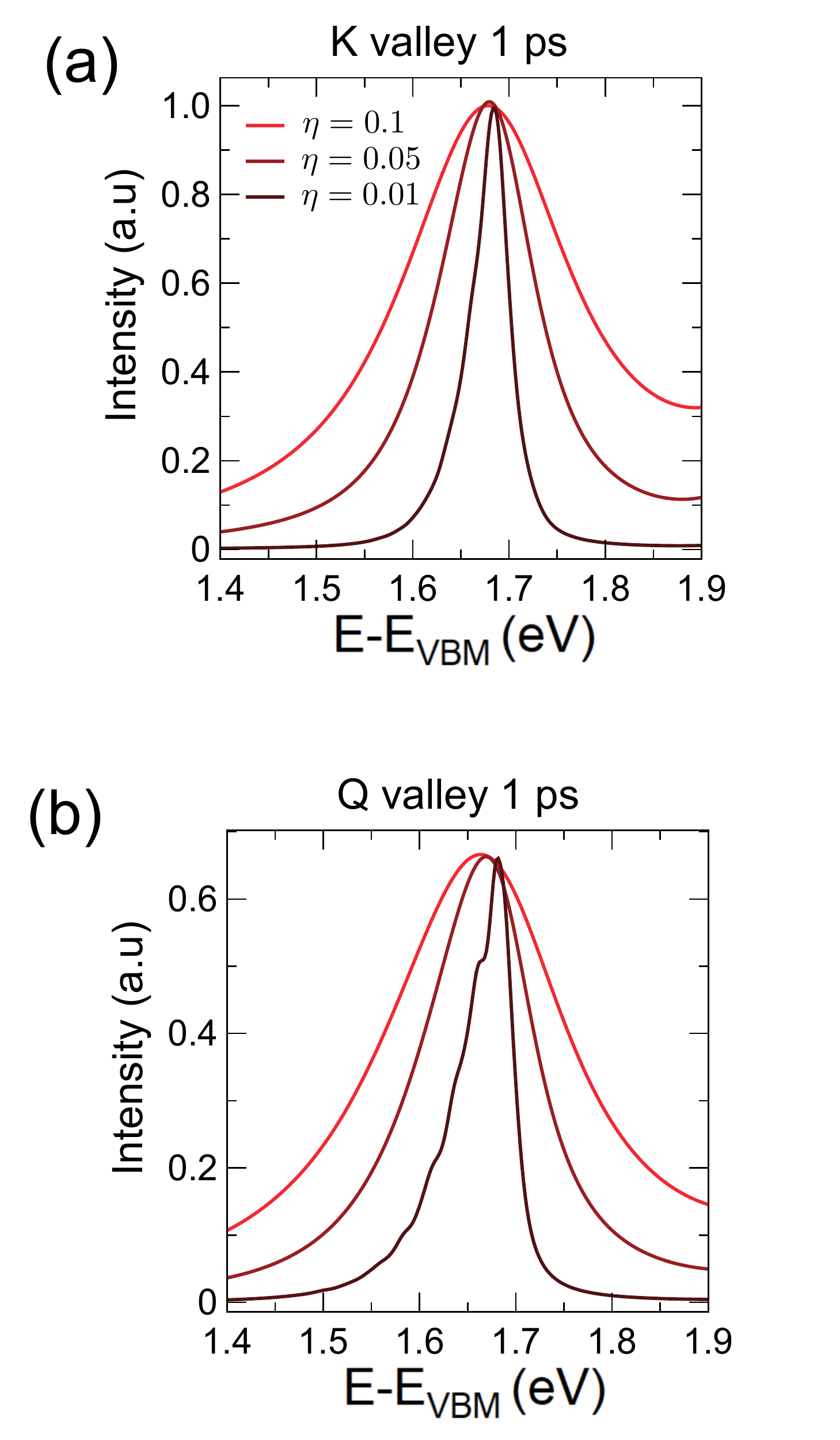}
\caption{{\bf$|$ Intrinsic nature of lineshape.}      
(h-i) Theoretical analysis of the asymmetry in the EDCs at $\tau=1$~ps for the
 $K$ and $Q$ valleys as a function of broadening parameter 
 $\eta$ varied between 0.01~eV and 0.1~eV. 
 This analysis  provides further evidence that 
the asymmetry of the EDC's is   an 
intrinsic feature, and not an artifact of the experimental resolution.
By reducing the broadening parameter  
the tail toward lower energies becomes increasingly more evident.
}
\label{EDfig3}
\end{extendedfigure}

\newpage

\begin{extendedfigure}[H]
\centering
\includegraphics[width=8cm, clip]{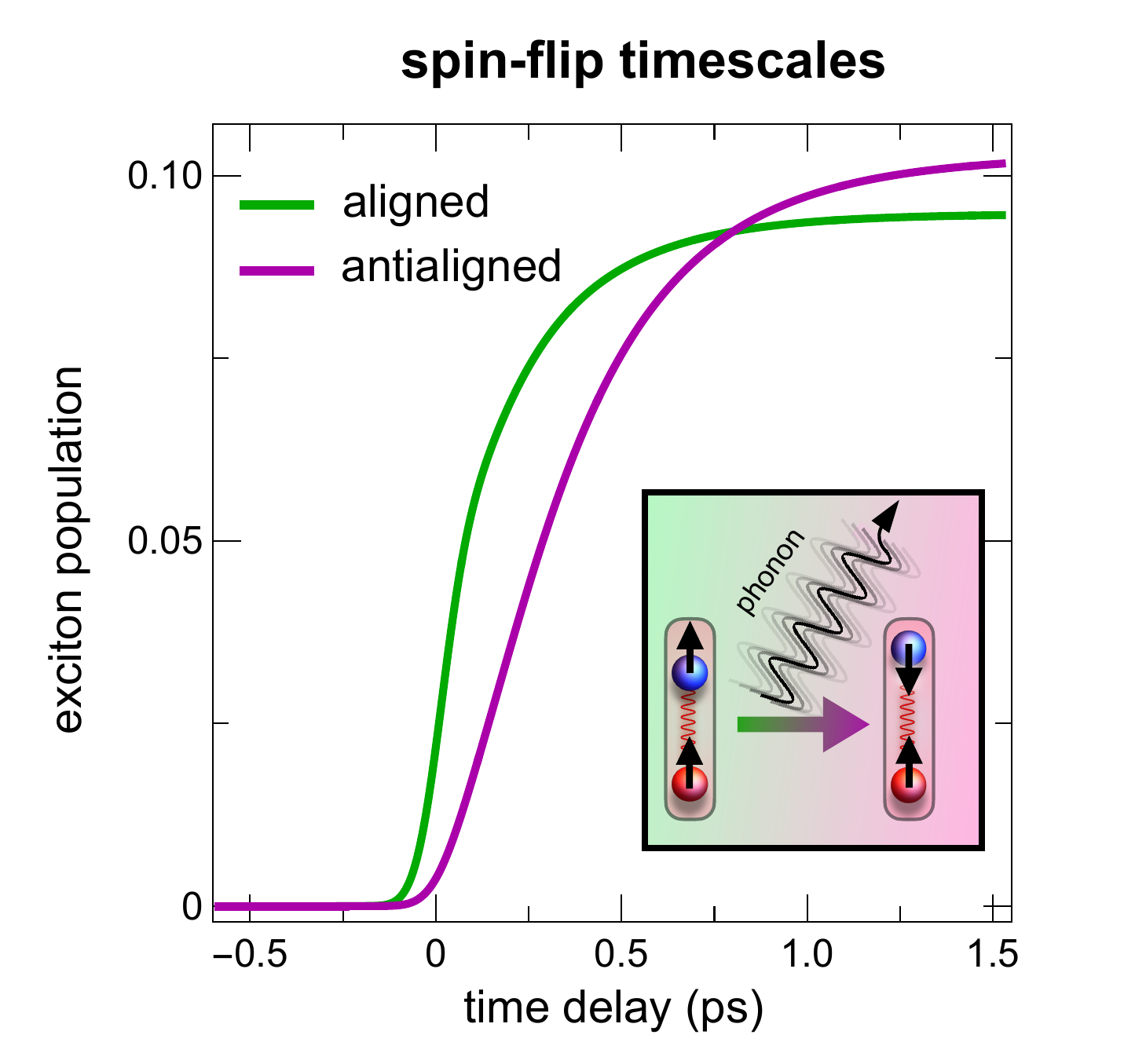}
\caption{{\bf $|$ Spin flip 
scattering.}                  
Simulated temporal evolution of the total population of spin-aligned (green curve)
and spin-antialigned (violet curve) bound excitons.
As the system evolves, excitons accumulate around the high-symmetry points. 
According to our Bethe–Salpeter equation (BSE) analysis, and 
consistent with previous reports~\cite{Deilmann_2019,PhysRevResearch.4.043203}, 
the lowest-energy excitons around
$\rm{K}$ and $\rm{Q}$ exhibit spin-aligned character
while those around $\Gamma$ and $\rm{M}$ are spin-antialigned.
Since photoexcitation only generates spin-aligned electron–hole pairs,
the formation of spin-antialigned 
excitons requires spin-flip processes, which 
in WSe$_{2}$ are primarly mediated by the emission 
of $E''$ optical phonons~\cite{Zhang_2022}. 
 As a result, the population of spin-antialigned excitons
(mainly from $\Gamma$ and $\mathrm{M}$) emerges at later
times and grows more slowly than that of spin-aligned excitons (mainly
from $\mathrm{K}$ and $\mathrm{Q}$). Nevertheless,
the populations of the two spin species become comparable within 
$\sim$1~ps.
 }
\label{SIfig1}
\end{extendedfigure}


\bibliography{sn-bibliography}

\end{document}